# evolSOM: an R Package for evolutionary conservation analysis with SOMs

Santiago Prochetto[1,2], Renata Reinheimer[3], Georgina Stegmayer[2]

[1]Instituto de Agrobiotecnología del Litoral, Universidad Nacional del Litoral, CONICET, CCT-Santa Fe, Ruta Nacional N° 168 Km 0, s/n, Paraje el Pozo, Santa Fe, Argentina.

[2]Research Institute for Signals, Systems and Computational Intelligence, sinc(i), FICH-UNL, CONICET, CCT-Santa Fe, Ruta Nacional N° 168 Km 0, s/n, Paraje el Pozo, Santa Fe, Argentina.

[3]Instituto de Agrobiotecnología del Litoral, FCA-UNL, CONICET, CCT-Santa Fe, Ruta Nacional N° 168 Km 0, s/n, Paraje el Pozo, Santa Fe, Argentina.

**Abstract**

**Motivation:** Unraveling the connection between genes and traits is crucial for solving many biological puzzles. In the intricate structure of DNA, genes provide instructions for building cellular machinery, directing the processes that sustain life. RNA molecules and proteins, derived from these genetic instructions, play crucial roles in shaping cell structures, influencing reactions, and guiding behavior. This fundamental biological principle links genetic makeup to observable traits, but integrating and extracting meaningful relationships from this complex, multimodal data presents a significant challenge.

**Results**: We introduce evolSOM, a novel R package that utilizes Self-Organizing Maps (SOMs) to explore and visualize the conservation of biological variables, easing the integration of phenotypic and genotypic attributes. The package enables the projection of multi-dimensional expression profiles onto easily interpretable two-dimensional grids, aiding in the identification of conserved gene modules/phenotypes across multiple species or conditions. By constructing species-specific or condition-specific SOMs that capture non-redundant patterns, evolSOM allows the analysis of displacement of biological variables between species or conditions. Variables displaced together suggest membership in the same regulatory network, and the nature of the displacement (early/delayed/flip) may hold biological significance. These displacements are automatically calculated and graphically presented by the package, enabling efficient comparison and revealing both conserved and displaced variables. The package facilitates the integration of diverse phenotypic data types, such as morphological data or metabolomics, enabling the exploration of potential gene drivers underlying observed phenotypic changes. Its user-friendly interface and visualization capabilities enhance the accessibility of complex network analyses. As an illustrative example, we employed evolSOM to study the displacement of genes and phenotypic traits, successfully identifying potential drivers of phenotypic differentiation in grass leaves.

**Availability:** The package is open-source under the GPL (>= 3) and is available at https://github.com/sanprochetto/evolSOM. A vignette with a step-by-step process is also available.

**Contact**: sprochetto@gmail.com

# 1 Introduction

In recent years, there has been a notable surge in technological advancements in the life sciences field. Researchers now have access to extensive biobanks and can collect vast amounts of multimodal data, including data from omics studies across multiple organisms, which is crucial for understanding evolution. Central to this understanding is the unraveling of the connection between genes and traits, an essential endeavor for solving various biological puzzles. Genes serve as instructions for building cellular machinery directing processes that sustain life. RNA molecules and proteins, products of these genetic instructions, play pivotal roles in shaping cell structures, influencing reactions, and guiding behavior. While this fundamental biological principle links genetic makeup to observable traits, the challenge lies in integrating and extracting meaningful relationships from this complex, multimodal data.

One of the key challenges of this new era is the development of robust tools capable of integrating diverse data types like biological variables (f.e. genes, phenotypes, among others) to uncover hidden patterns and relationships among heterogeneous sources. Self-organizing maps (SOM) have gained interest among scientists due to their ability to integrate and visualize heterogeneous data through unsupervised learning (Stegmayer *et al*., 2012; Milone *et al*., 2013; Watanabe and Hoefgen., 2019, Betts *et al*., 2020; Mohnike *et al*., 2023). They can represent complex high-dimensional input patterns into a simpler low-dimensional discrete map, with prototype vectors (neurons) that can be visualized in a two-dimensional lattice structure and which preserve the proximity relationships of the original samples (Kohonen *et al*., 2001). By examining each data point individually, and clustering variables with highly similar expression patterns along measurements, SOMs enable an individualized analysis of each biological variable rather than a global one. In particular, SOMs excel at capturing non-linear relationships and mapping data onto visually intuitive grids, allowing researchers to not only identify general trends but also explore particular differences within individual variables. So far, a single package that streamlines this process in a straightforward manner is currently unavailable.

In this application note, we present evolSOM, a novel tool based on unsupervised machine learning that facilitates the discovery of variations between multimodal data and offers an intuitive way to visualize hidden patterns within complex data matrices. The package facilitates the integration of diverse phenotypic data types, such as morphological data or metabolomics, enabling the exploration of potential gene drivers underlying observed phenotypic changes. Specifically, evolSOM clusters biological variables (genes and/or phenotypic traits) based on their expression patterns along several conditions and maps these variables onto a reference SOM, allowing the analysis of displacement between species.

Variables displaced together suggest membership in the same regulatory network, and the nature of the displacement itself (early/delayed/flip) may hold biological significance. These displacements are automatically calculated and graphically presented by the package, enabling efficient comparison and revealing both conserved and displaced biological variables. Its user-friendly interface and visualization capabilities enhance the accessibility of complex network analyses. As an illustrative example, we employed evolSOM to study the displacement of genes and phenotypic traits. This analysis successfully identified potential drivers of phenotypic differentiation in grass leaves (Prochetto *et al.*, 2024).

## 2 The evolSOM R package

The package analyzes multi-modal data through a series of automated steps outlined in Figure 1. Initially, pre-processed data, such as gene counts, protein levels, and anatomical measures, are imported and scaled to accommodate diverse data types. Following this, a "reference" Self-Organizing Map (SOM) is constructed for a control condition, optimizing the map size for efficiency. Subsequently, biological variables from additional "test" species or conditions are mapped onto the reference SOM, allowing for the assessment of conservation and displacement of the variables based on their respective neuron locations. The package then automatically identifies the nature of these displacements (early/delayed/flip) and quantifies their overall occurrence. Finally, it generates visualizations to emphasize the relative abundance of biological variables on the map and provides detailed insights into their specific displacements. The following sections explain in detail every step of the workflow.

### 2.1 Data import and scaling

The package accepts data in comma-separated value (CSV) format. Rows represent biological variables (e.g., gene transcripts, phenotypes, metabolites), while columns represent measured conditions (e.g., tissues, cells, developmental stages, species). Data from different species could be integrated into one dataframe or split into one dataframe per species. An additional dataframe can be included to provide supplementary information about each variable, such as its type (gene or phenotypic trait), involvement in biological processes, gene family affiliation, and more.

Scaling the data before constructing a SOM constitutes a critical preprocessing step that enhances the performance, interpretability, and robustness of the SOM. If variables have different scales, those with larger scales dominate distance calculations, compromising accurate clustering. Scaling ensures all variables contribute equally to similarity calculations and enhances robustness to outliers. The **scale_species()** function scales data to unit variance for each species or condition separately, returning a list containing one data frame per species.

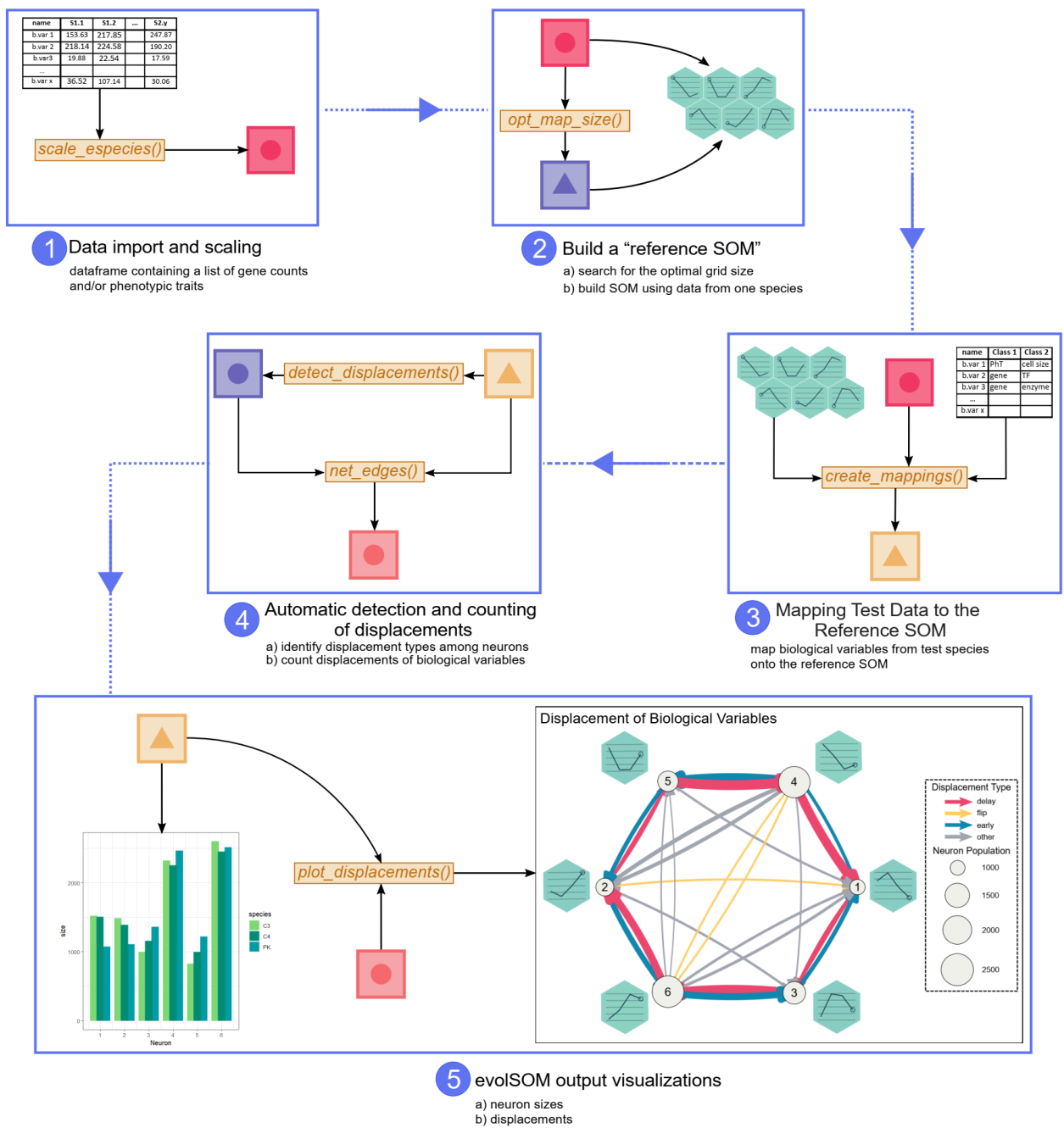

**Figure 1**: Workflow using evolSOM R package.

## 2.2 Building a "reference SOM"

In order to illustrate the conservation and displacement of biological variables across diverse species or conditions, the evolSOM package constructs a "reference SOM." However, determining its optimal grid size is crucial. This optimal size ensures the faithful representation of all expression patterns by neurons, ensuring each pattern's uniqueness. Manual determination of the best grid size can be challenging, particularly for those less familiar with SOMs. The optimal grid size depends not only on the number of biological variables under analysis but also on the conditions under which the data was measured.

The **opt_map_size()** algorithm systematically evaluates different grid sizes to encompass all non-redundant expression patterns. The resulting "map_size" defines the grid's optimal dimensions. Illustrated in Supplementary Material 1, this algorithm takes inputs such as the dataset *D*, the correlation threshold $\vartheta$, the initial map size *d*, and the number of SOM iterations *i*. The initial map size s is determined by the formula $d^2 + 1$, and the variable *X* is initialized to count neurons with correlations exceeding the threshold. The SOM dimensions *d1* and *d2* are then determined at the ceiling and floor of s-1, and multiple SOMs are iteratively built with dimensions *d1* and *d2*, trained with dataset *D*. For each trained map, pairwise correlations $\varrho_{i,j}$ of all map neurons are calculated and checked against the correlation threshold $\vartheta$. The number of neurons exceeding $\vartheta$ is recorded. A null count indicates that all neurons are above the correlation threshold, ensuring an optimal distribution of data samples across the map. When this condition is fulfilled, the optimum SOM map dimensions are returned. With the determined optimal SOM map dimensions, the reference SOM is constructed and visualized using the aweSOM package (v1.3, Boelaert *et al*., 2022).

### 2.3 Mapping Test Data to the Reference SOM

After building the "reference SOM" using the data from the control species, the data from the test species is mapped into the "reference SOM" to calculate displacements. The **create_mappings()** function is employed for this purpose. Along with the test data, the inputs include the reference SOM and the data used for its construction. Additionally, an extra dataframe is incorporated to provide supplementary categorical information about each variable. The mappings object stores valuable information about the mappings, such as class assignments and number of features allocated to each neuron. The results include a "class dataframe", indicating the original neuron location in the reference SOM for each biological variable, along with its new location after mapping.

### 2.4 Automatic detection and counting of displacements

To uncover hidden evolutionary patterns between the test species and the reference species, the package employs the **detect_displacements()** function to identify movements or displacements among conditions. This function scrutinizes relationships between expression patterns of neurons in the reference SOM, categorizing these relationships into four types: early, delay, flip, and other.

The discovery of displacements involves, essentially, the cross-correlation between two data series. If the cross-correlation is maximum at the middle point of the data series, no displacement is observed. When the correlation is maximum at the right part of the middle point and surpasses the delay_threshold, it indicates a delay displacement. Similarly, if the correlation is maximum at the left part of the middle point, it indicates an early displacement. Conversely, a negative cross-correlation lower than the flip_threshold indicates a flip. The displacement object provides information about the detected displacement types, facilitating a comprehensive exploration of evolutionary displacement patterns between species.

Upon characterizing relationships between neurons, the **net_edges()** function creates a dataframe detailing the number of displacements between neurons and species pairwise. This output facilitates analysis of prevalent displacement types, or identification of broader trends, among others.

**2.5 evolSOM output visualizations**

Using information from the **create_mapping()** output a bar graph can be plotted to illustrate the allocation of biological variables to each neuron in both the reference SOM and the test conditions.

The output from **create_mapping()** and **net_edges()** is instrumental in plotting a displacement graph using the **plot_displacements()** function. This visualization represents neurons as circles, with sizes corresponding to the number of variables allocated to them. Displacements are depicted as arrows connecting neurons, the lengths of which are proportional to the number of variables displaced. Additionally, by selecting a specific group of biological variables, researchers can explore whether a distinct displacement pattern exists for this particular group. This offers valuable insights into the evolution of this group of variables and their relationships between the reference species and the test species.

**3 $C_4$ Photosynthesis as a Case Study**

To exemplify evolSOM's power in dissecting gene-trait relationships, we revisit a prior study (Prochetto *et al*., 2024) focusing on $C_4$ photosynthesis, a highly efficient carbon fixation pathway found in various plant lineages. In that work, we employed evolSOM to shed light on leaf anatomy differentiation and evolution in three non-model grass species with different photosynthetic pathways ($C_3$, PK, and $C_4$). Combining leaf transcriptome data with detailed anatomical measurements, we aimed to identify the genetic underpinnings of this lineage leaf structural diversity.

We started by integrating transcriptomic data from a previous study with anatomical information from four leaf segments representing different developmental stages (Prochetto *et al*., 2023). After scaling the data, an optimal map size of 2x3 was determined for our dataset comprising 9757 biological variables. The resulting map consisted of 6 neurons, each exhibiting distinct expression patterns. Following this, we constructed a reference SOM utilizing data from the $C_3$ species, and subsequently, we mapped the data from PK and $C_4$ onto this reference SOM for further analysis.

Our analysis centered on bundle sheath (BS) cell size, a crucial phenotypic trait in $C_4$ anatomy (Sage *et al*., 2014). Using evolSOM, we identified 41 genes with expression patterns intricately linked to BS cell size across $C_3$, PK, and $C_4$ grass species. Among the displaced genes were two known components of suberin biosynthesis, a characteristic polymer in $C_4$ bundle sheath cell walls. Additionally, our analysis identified two displaced transcription factors, suggesting potential regulatory mechanisms underlying the observed phenotypic shift. This case study demonstrates evolSOM's ability to unveil expression displacement patterns linked to adaptation and identify key biological processes and candidate regulatory genes associated with complex evolutionary transitions.

## 4 Conclusions

Our package represents a valuable contribution to biologists by offering a powerful toolset for analyzing, visualizing, and interpreting complex multi-modal data. By integrating genotypic and phenotypic attributes, evolSOM provides a valuable resource for unveiling intricate relationships within biological systems. Its distinct ability to map data onto Self-Organizing Maps (SOMs) and analyze displacement patterns facilitates a nuanced exploration of evolutionary dynamics across species or conditions.

Beyond its primary applications outlined in the preceding sections, the evolSOM package holds versatile potential for other scientific investigations. The package could find utility in the exploration of temporal dynamics within biological systems, enabling the identification of temporal shifts in gene expression or phenotypic traits. Similarly, it can be applied to investigate spatial dynamics within organisms or analyze knockdown or overexpression lines alongside wild-type data in genetic experiments. Furthermore, evolSOM may contribute to understanding inter-individual variability by analyzing and comparing diverse datasets, such as omics data from individuals.

Its user-friendly interface and visualization capabilities make it accessible to a broad audience, extending its utility beyond bioinformatics experts to researchers across various domains interested in exploring complex relationships within multi-dimensional datasets.

**Conflict of Interest**

None declared.

**Supplementary data**

Supplementary Material 1: Algorithm for the determination of the optimum map size.

**Author contributions**

Conceptualization: SP, GS and RR; Data curation: SP; Funding acquisition: GS and RR; Methodology, Software, Visualization and Writing - original draft : SP and GS; Supervision: GS; Writing – review & editing: SP, GS and RR